\documentclass[prl,aps,twocolumn,floatfix,amsmath,amssymb,superscriptaddress,tightenlines]{revtex4}
\usepackage{graphicx}
\usepackage{amsfonts}
\usepackage{bm}
\usepackage{color}
\usepackage{ulem}
\begin{document}

\date{\today}
\title{Measuring Renyi Entanglement Entropy with Quantum Monte Carlo}

\author{Matthew B. Hastings}
\affiliation{Microsoft Research, Station Q, CNSI Building, University of California, Santa Barbara, CA, 93106}

\author{Iv\'an Gonz\'alez}
\affiliation{Centro de Supercomputaci\'on de Galicia, Avda.~de~Vigo~s/n, E-15705 Santiago de Compostela, Spain}

\author{Ann B. Kallin}
\affiliation{Department of Physics and Astronomy, University of Waterloo, Ontario, N2L 3G1, Canada} 

\author{Roger G. Melko}
\affiliation{Department of Physics and Astronomy, University of Waterloo, Ontario, N2L 3G1, Canada} 

\begin{abstract}

We develop a quantum Monte Carlo procedure, in the valence bond basis, to measure the Renyi entanglement entropy of a
many-body ground state as the expectation value of a unitary {\it Swap} operator acting on two copies of the system.
An improved estimator involving the ratio of {\it Swap} operators for different subregions enables convergence of the entropy in a simulation time polynomial in the system size.  We demonstrate convergence of the Renyi entropy to exact results for a Heisenberg chain. Finally, we calculate the scaling of the Renyi entropy in the two-dimensional Heisenberg model and confirm that the N\'eel groundstate obeys the expected area law for systems up to linear size $L=32$.

\end{abstract}
\maketitle

The measurement of entanglement entropy  provides
new tools to determine universal properties of interacting quantum many-body systems in condensed-matter physics.  For example, conformally invariant
systems in one dimension (1D) display universal logarithmic
scaling of entanglement entropy \cite{Cardy}.  Many
quantum systems in two dimensions (2D) and higher are predicted to have ``area law''
scaling of entanglement entropy, with
either universal additive logarithmic corrections for critical systems \cite{corner,ryu} or
universal additive constant corrections for topologically ordered systems \cite{KP,LW}.

Scalable simulation methods have become crucial for the 
study of groundstate, finite-temperature, and critical properties of quantum
many-body systems.
Unfortunately, there are 
no known scalable simulation methods to calculate entanglement entropy in $D>1$.
Calculating the von Neumann entropy ($S_1$, defined below) requires calculating
the reduced density matrix,
which can be done in Lanczos 
or density matrix renormalization group (DMRG).  However, scalable simulation techniques in dimensions $D>1$,
are so far essentially
restricted to quantum Monte Carlo (QMC), which does not provide access to the reduced density matrix, but only approximates it
via a Metropolis importance-sampling algorithm. 

We resolve this problem and show that a generalized Renyi entropy ($S_2$, defined below) can be related to the expectation
value of a unitary {\it Swap} operator in the valence bond basis,
directly accessible in QMC.
Using QMC simulations, we calculate $S_2$ for a spin $1/2$ Heisenberg model
on 1D and 2D lattices.  In 1D, we demonstrate that $S_2$ calculated with QMC converges to the exact result obtained with DMRG simulations.  In 2D, we show that the leading-order term in the scaling of $S_2$ goes like the boundary of a subregion $A$, 
confirming the area law expected from recent studies \cite{Ann}.

\begin{figure} {
\includegraphics[width=2.4 in]{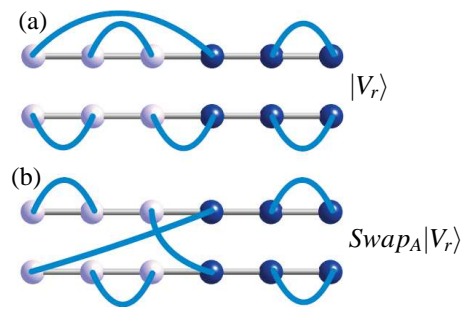} \caption{(color online) 
\label{swap_2}
A six-site chain, with two non-interacting copies (top and bottom) before
(a) and after (b) the $Swap_A$ operation. 
The region $A$ consists of three light-colored sites on the left; the
complement region $B$ of three dark sites on the right.  The curved lines denote
singlets in the state $|V_{r}\rangle$, which is a product of two
different valence bond states, one per copy.
The ground state of the entire system is a linear combination of similar $|V_{r}\rangle$.
}
} \end{figure}

The generalized Renyi entanglement entropies are defined by
\begin{equation}
S_{n} (\rho_A) = \frac{1}{1- n} \ln \left[{ {\rm Tr}\big( \rho_A^{n} \big) }\right],
\label{Sn}
\end{equation}
where $\rho_A$ is the reduced-density matrix of a subregion $A$ entangled with its complement $B$, and 
$n>0$.  In the limit $n \rightarrow 1$, one recovers the 
familiar von Neumann entropy, $S_1 (\rho_A) = - {\rm Tr}( \rho_A \ln \rho_A )$.  
Recently, the generalized Renyi entropies have attracted considerable attention in the condensed matter community, due to their ability to encode
information about the whole ``entanglement spectrum'' of $\rho_A$, together allowing the set of $S_{n}$ to contain 
much more information than $S_1$ alone \cite{Espec}.  For example, the concept of topological entanglement entropy has recently been
generalized to the family of Renyi entropies, where it has been shown to be equal to the logarithm of the total quantum 
dimension, $S_{\rm top} \propto \log {\mathcal D}$, independent of $n$ \cite{PI}.  Universal corrections to the scaling of $S_{n}$ 
have also been calculated in a field-theoretic treatment of the $O(N)$ model \cite{Max}.
Further, any two Renyi entropies $S_{n}$ and $S_{m}$ obey the
inequality $S_{n} \geq S_{m}$ for $n < m$, making $S_2$ a
useful lower bound on $S_1$.  

{\it Valence bond basis and the Swap operator.}-- We briefly review the notation of Sandvik, and refer the reader to 
Refs.~\cite{Sandvik,Beach,AWSloop} for additional details on the valence bond basis.
One begins by writing the expansion of a singlet
wavefunction $\Psi_0$ of $N$ spins as $N/2$ valence bonds:
\begin{equation}
| \Psi_0 \rangle = \sum_r f_r|(a^r_1,b^r_1) \ldots (a^r_{N/2},b^r_{N/2}) \rangle = \sum_r f_r| V_r \rangle
\end{equation}
where the valence bonds are singlet pairs denoted by $(a,b) = (\lvert\uparrow_a \downarrow_b \rangle - \lvert\downarrow_a \uparrow_b\rangle)/\sqrt{2}$,
occurring between sites $a$ and $b$ of the opposite sublattices of a bipartite lattice, and $r$ labels all possible tilings of bonds.  The coefficients $f_r$ are unknown, but an importance sampling scheme \cite{Sandvik} for the valence bond state $| V_r \rangle$ will be outlined below.  

%
%
To derive an estimator for $S_2$, we construct two copies of the
system (higher entropies $S_n$ require $n$ copies);
one ``real'', one ``ancillary'' (Fig.~\ref{swap_2}). Then we define an
operator $Swap_A$ which acts between the copies of the system, swapping
the configurations within region $A$.  To define this operator, we
give its matrix elements in a product basis, e.g.  the
$S^z$ basis.  Let $|\alpha\rangle$  be a complete basis of states in the
region $A$ and let $|\beta\rangle$ be a complete basis of states in the
complement region $B$. The state of each copy can be decomposed
in this product basis as $|\Psi\rangle=\sum_{\alpha,\beta} C_{\alpha,\beta}
|\alpha\rangle | \beta \rangle$ for some amplitudes $C_{\alpha,\beta}$.
We define the unitary operator $Swap_A$ by
\begin{eqnarray}
\label{SwapDef}
&& Swap_A  \Bigl( \sum_{\alpha_1,\beta_1} C_{\alpha_1,\beta_1}|\alpha_1\rangle |\beta_1\rangle \Bigr) \otimes
\Bigl( \sum_{\alpha_2,\beta_2} D_{\alpha_2,\beta_2}|\alpha_2\rangle
|\beta_2\rangle \Bigr) \nonumber\\ &=&
\sum_{\alpha_1,\beta_1} C_{\alpha_1,\beta_1}
\sum_{\alpha_2,\beta_2} D_{\alpha_2,\beta_2}
\Bigl( |\alpha_2\rangle |\beta_1\rangle \Bigr) \otimes
\Bigl( |\alpha_1\rangle |\beta_2\rangle \Bigr).
\end{eqnarray}
The two copies of the system will not interact with each other, so the
ground state of the combined system will be the product of the ground states
on each copy:
$|\Psi_0\otimes \Psi_0\rangle$.  Thus, the expectation value
of $Swap_A$ is
\begin{align}
\label{H2fromSwap}
 \langle \Psi_0\otimes \Psi_0| & Swap_A|\Psi_0\otimes \Psi_0\rangle
\nonumber \\ 
&= 
\sum_{\alpha_1,\alpha_2,\beta_1,\beta_2} C_{\alpha_1,\beta_1} \overline C_{\alpha_2,\beta_1} C_{\alpha_2,\beta_2} \overline C_{\alpha_1,\beta_2}
\nonumber \\
&=
\sum_{\alpha_1,\alpha_2} (\rho_A)_{\alpha_1,\alpha_2}
(\rho_A)_{\alpha_2,\alpha_1}={\rm Tr}(\rho_A^2),
\end{align}
where $(\rho_A)_{\alpha_1,\alpha_2}=\sum_{\beta_1}C_{\alpha_1,\beta_1} \overline C_{\alpha_2,\beta_1}$
denotes a matrix element of $\rho_A$. Therefore, 
\begin{equation} 
\label{s2}
S_2(\rho_A) =-\ln({\rm Tr}(\rho_A^2))=-\ln(\langle Swap_A \rangle). 
\end{equation}
Equation~(\ref{s2}) is basis independent~\cite{Jaeger}. In particular it
holds in the valence bond basis, where the $Swap_A$ operator is not
defined by Eq.~(\ref{SwapDef}). Rather, it is defined to swap the
endpoints of valence bonds between the real and ancillary copies in the
region $A$, as illustrated in Fig.~\ref{swap_2}.  

{\it Measuring the Swap operator in QMC.}--
We now present a procedure for measuring the entropy $S_2$ via importance sampling of the $Swap_A$ operator in QMC.  A
projector scheme in the valence bond basis has recently been pioneered by Sandvik, and we only briefly
review the notation here, referring the reader to Refs.~\cite{Sandvik,Beach,AWSloop} for details of implementation.
The method is a $T=0$ projector QMC, where high powers of
a Hamiltonian $H$ are used to project out the ground state $|\Psi_0\rangle$ from a trial wavefunction $\Psi$:
$(-H)^m|\Psi \rangle \propto |\Psi_0\rangle$.  In Sandvik's QMC scheme, $(-H)^m$ is written as a sum of all products of $m$ bond
operators,
\begin{equation}
(-H)^m = \sum_r \prod_{i=1}^m H_{a_i^r b_i^r} = \sum_r P_r,
\end{equation}
where for concreteness we use the spin 1/2 Heisenberg model, defining $H=-\sum_{\langle a,b \rangle}H_{ab}$ and 
$H_{ab} = -({\bf S}_a \cdot {\bf S}_b - 1/4)$.  The ``operator string'' $P_r$ is sampled according to its weight, $W_r$, accrued upon
evolution of a trial valence bond state $|V \rangle$ under projection:
\begin{equation}
P_r |V \rangle = W_r |V(r) \rangle.
\end{equation} 
As shown in Ref.~\cite{Sandvik}, for the Heisenberg model, $W_r$ is simply related to the number of off-diagonal 
operations $m_{\rm off}$ in the projection, $W_r = 2^{-m_{\rm off}}$.

To sample the $Swap_A$ operator, one requires a double-projector valence bond QMC scheme \cite{Sandvik}, where
a general expectation value for any observable $\mathcal{O}$ is given by
\begin{equation}
\label{expect}
\langle \mathcal{O} \rangle = \frac{\sum_{rl} \langle V | P_l^* \mathcal{O} P_r | V \rangle} {\sum_{rl} \langle V | P_l^* P_r | V \rangle} 
= \frac{\sum_{rl} W_l W_r \langle V(l) | \mathcal{O} | V(r) \rangle} {\sum_{rl} W_l W_r \langle V(l) | V(r) \rangle}.
\end{equation} 
In this case, the two operator strings, $P_l$ and $P_r$, are applied to two copies of the system; 
the expectation value of the $Swap_A$ operator as illustrated in Fig.~\ref{swap_2} 
can then be calculated directly.  Specifically, one 
performs importance sampling of operator strings according to the weight $W_l W_r \langle V(l) | V(r) \rangle$, and measures
the QMC average expectation value
\begin{equation}
\langle Swap_A \rangle =  \left\langle{ \frac{ \langle V(l) | Swap_A | V(r) \rangle}{\langle V(l) | V(r) \rangle}  }\right\rangle,
\label{Swap}
\end{equation}
calculating $S_2$ from Eq.~\eqref{s2}.

\begin{figure} {
\includegraphics[width=2.9in]{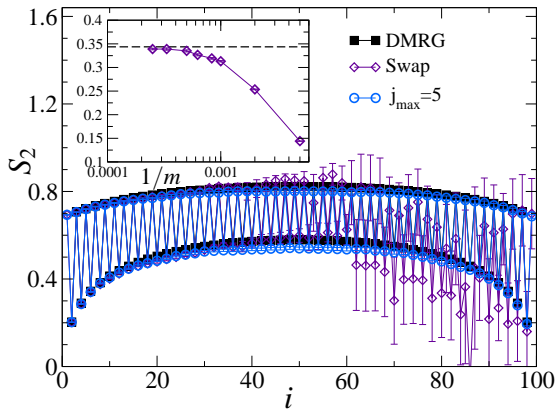} \caption{(color online) 
\label{1Dfig}
The Renyi entropy $S_2$ as a function of site index $i \in A$, for a 100-site Heisenberg chain with open boundaries, 
calculated with DMRG and QMC.  Data labeled ``Swap'' was calculated with Eq.~\eqref{Swap} with one QMC simulation, while
data labeled $j_{\rm max}=5$ was calculated with Eq.~\eqref{Ratio} using 20 separate QMC simulations with a range of  $j \in [1,5]$.  The inset shows the convergence of $S_2$ to the exact value (dashed line) for $i=6$ with up to $m=4000$.
}
} \end{figure}

Results for a 1D chain are illustrated in Fig.~\ref{1Dfig}.  There, simulations were performed using a double-projector QMC,  with a
simple ``columnar'' trial state $| V \rangle$ (alternating nearest-neighbor bonds in 1D).  In the following data, four random bond operators were changed in each operator string per Monte Carlo step, and (unless otherwise
stated) $m=40N$.  Exact results for $S_2$ were obtained with a DMRG simulation, converged with 1000 states.  
One 
immediately sees that the naive expectation value $\langle Swap_A \rangle$ 
results in very large statistical error bars 
when the region
$A$ grows large (in 1D, $A$ is the linear region of size $i$).  
One may understand this by considering the expected scaling of $S_2$.  Namely,
$S_2$ scales logarithmically with the size of $A$, so the expectation value
$\langle Swap_A \rangle$ should be polynomially small in $A$.  As a result, there
are only very
few configurations contributing significantly to $\langle Swap_A \rangle$, so in practice the importance sampling is done poorly,
resulting in large error bars and possibly jeopardizing simulation ergodicity \footnote{The valence bond basis does alleviate this problem, compared to the
$S^z$ basis, in some systems; for example, a valence
bond crystal can have an exponentially small $\langle Swap_A \rangle$, without having {\it any} realization-to-realization fluctuations,
while in the $S^z$ basis the expectation value  in a given realization is either $0$ or $1$.}.
The problem is even worse in 2D and higher, where the expectation value 
$\langle Swap_A \rangle$ should be exponentially small in $A$.

To combat this issue, we propose a refinement to the algorithm, which we called ``improved ratio'' sampling.
Consider first a one-dimensional chain, and define regions $A^1,A^2,...,A^n$, such
that $A^i$ contains $i$ sites, and $A^i$ is a subset of $A^{i+1}$.  In other words, each region $A^{i+1}$ is obtained by adding one site to region
$A^{i}$, and region $A^0$ is the empty set (thus, $Swap_{A^0}$ is equal to the identity operator).
In a given simulation, one can calculate the ratio
\begin{equation}
\frac{\langle Swap_{A^{i+1}}\rangle}{\langle Swap_{A^{i}}\rangle}
= \frac{\sum_{rl} W_l W_r \langle V(l) | Swap_{A^{i+1}} | V(r) \rangle} {\sum_{rl} W_l W_r \langle V(l) | Swap_{A^i} | V(r) \rangle}
\label{Ratio}
\end{equation}
where, for each $i=0,...,n-1$, the log of this ratio is equal to minus the difference $S_2(\rho_{A^{i+1}})-S_2(\rho_{A^{i}})$.
Computing this ratio for each $i$ will let us compute $S_2$ for all $A^i$.  To calculate this ratio, the simulation must be performed 
with the modified sampling weight,
\begin{equation}
W_l W_r \langle V(l) | Swap_{A^i} | V(r) \rangle,
\end{equation}
i.e., a unique QMC simulation must be done for each desired $i$.  
If every site were used for a different simulation weight, the QMC algorithm would obtain an additional multiplicative factor of $N$ in its scaling.
In practice, 
this can be reduced by noting that good statistical control can be retained by calculating 
${\langle Swap_{A^{i+j}}\rangle}/{\langle Swap_{A^{i}}\rangle}$ for fixed $i$ and a range of $j \in [1,j_{\max}]$ (see also the Discussion).  We illustrate this in Fig.~\ref{1Dfig},
where simulations with $j_{\rm max}=5$ are sufficient to converge 100-site chain to the exact results along most of its length.

{\it 2D Heisenberg Results.}-- We now extend these concepts to QMC simulations of the spin 1/2 Heisenberg model on $N=L \times L$ lattices.  
For the simulations using the improved ratio estimator, 
${\langle Swap_{A^{\ell+r}}\rangle}/{\langle Swap_{A^{\ell}}\rangle}$,
we define $A^\ell$ as a square region of linear size $\ell$ containing $\ell \times \ell$ sites, so
$A^{\ell+r}$ contains $2\ell r+r^2$ more sites than $A^{\ell}$.

\begin{figure} {
\includegraphics[width=2.8in]{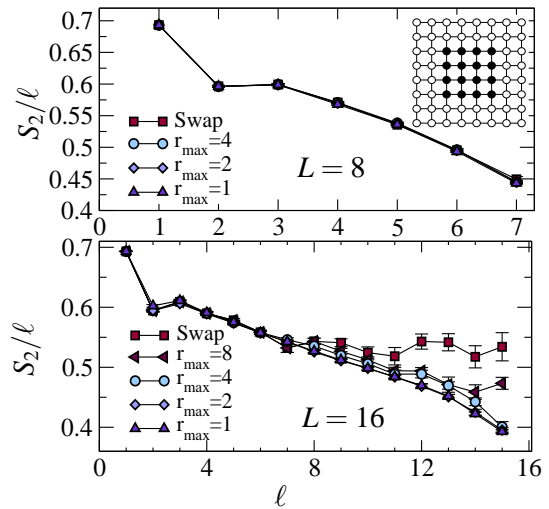} \caption{(color online) 
\label{L16fig}
The Renyi entropy divided by the linear dimension $\ell$ of the entangled region $A$, for 2D lattices with $L=8$ and $L=16$.  The data labeled ``Swap'' is from a single simulation calculating Eq.~\eqref{Swap} directly.  The other
data sets are derived from the improved ratio estimator, Eq.~\eqref{Ratio}, with different ranges of $r \in [1,r_{\rm max}]$ (see text).  The inset 
is a periodic $L=8$ lattice with region $A$ consisting of the 16 central (dark) sites labeled by $\ell = 4$.
}
} \end{figure}

Figure \ref{L16fig} illustrates $S_2$ on two 2D Heisenberg models with  $L=8$ and  $L=16$.  In the case of $L=8$, data for a single 
simulation calculating $\langle Swap_A \rangle$ directly is essentially identical to the improved ratio estimators for all $\ell$.  
Here, the direct estimator (Eq.~\eqref{Swap}) corresponds to one simulation, while $r_{\rm max}=1$ corresponds to 6 different simulations of the improved ratio estimator ${\langle Swap_{A^{\ell+1}}\rangle}/{\langle Swap_{A^{\ell}}\rangle}$ and $\ell \in [1,6]$.
In contrast to $L=8$, for $L=16$ the direct estimator is significantly different than 
the improved ratio estimator.  We can see the convergence of the data as one successively decreases the range of $r \in [1,r_{\rm max}]$, 
so that $r_{\rm max}=2$ is identical to $r_{\rm max}=1$, which is the smallest possible range for the improved ratio estimator in the 2D geometry illustrated.  It is important to note that for large $r_{\rm max}$ the correct value of $S_2$ does {\it not} lie within the error bars of the data.  
This is an indication that simulation ergodicity may be jeopardized by low sampling of the $Swap_A$ operator.
\begin{figure} {
\includegraphics[width=3.2in]{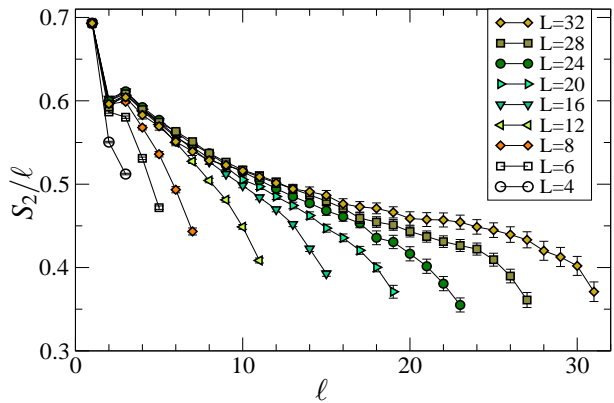} \caption{(color online) 
\label{fig4}
Scaling of the Renyi entropy divided by the linear dimension $\ell$ of the entangled region $A$ for different systems sizes
in 2D.  All data is calculated with the improved ratio estimator, Eq.~\eqref{Ratio}, with $r_{\rm max}=1$.  
}
} \end{figure}

We use the improved ratio estimator with $r_{\max}=1$ to scale the results for the Renyi entropy of the Heisenberg model to larger system sizes in 2D.  Results are given in Fig.~\ref{fig4} for $L=4$ to $L=32$, plotted as $S_2/\ell$.  From  Ref.~\cite{Ann}
one expects the scaling of $S_1$ in the N\'eel state to obey the area law; also recall that $S_2 \leq S_1$.  
The data illustrated is clearly consistent with the area law $S_2/\ell
\sim {\rm const.}$ for $\ell \ll L$, at which point boundary 
effects likely become important.  In particular, multiplicative log corrections (which were apparent in similar system sizes and geometries for the Valence-Bond entanglement entropy
\cite{Alet,Chh}) are not present.

{\it Discussion.}-- 
In this paper we have presented an algorithm for measuring the Renyi entropy $S_2$ in valence bond basis QMC simulations via
the expectation value of a swap operator between two copies of the system.
Using an improved ratio estimator, we are able to converge the expectation value of $S_2$ to the exact
result on a 1D Heisenberg chain.
Using the same procedure, we have presented the first measurement of the Renyi entanglement entropy in a 2D system, confirming the area law for the N\'eel groundstate of the Heisenberg model.

The simple double-projector QMC algorithm used in this paper is known to scale approximately as $O(m^2)$, \cite{AWSloop}
assuming $m >N$.
Thus, the current simulation results for the direct swap expectation
value, Eq.~\eqref{Swap}, also have
 $O(m^2)$ scaling, while results obtained using the improved ratio
estimator, Eq.~\eqref{Ratio}, have $O(Lm^2)$ scaling in the current geometries.
We note however that these geometries may not be ideal for very large $\ell$ and $L$, even in the improved ratio sampling.
For example, for $L=32$, results for $r_{\rm max}=2$ and $r_{\rm max}=1$, 
although remaining converged within error bars,
begin to develop larger discrepancies when $\ell$ approaches $L$.

From the current work 
(which used about 10 CPU-years), 
the question naturally arises whether it is possible to converge the expectation value of the swap operator on
larger system sizes. 
We expect that this will be possible using a different
sequence of regions $A^i$ in the improved ratio estimator.
Rather than considering a sequence of regions $A^{\ell}$ consisting of $\ell$-by-$\ell$ squares, 
we can use a more finely grained sequence of regions; e.g. by defining $A^i$ to contain
$i$ sites such that for $i=\ell^2$, $A^i$ is an $\ell$-by-$\ell$ square.  Incrementing the number of sites in $i$ by one would result in
(at worst) $O(Nm^2)$ scaling in the current algorithm.  
However, we also note that recently-developed {\it loop} algorithms
result in significant improvement in scaling, allowing for the convergence of observables such as energy and spin correlation functions for $L=256$ and larger \cite{AWSloop}.  In these schemes, the scaling of the direct swap operator (Eq.~\eqref{Swap}) 
would decrease to $O(m)$, while the expected scaling of the improved ratio estimator would be $O(Nm)$ at worst, depending
on the definitions of the $A^i$ regions.
We therefore expect that the algorithm will require only a polynomial number of samples to converge the Renyi entropy on arbitrary system sizes.

{\it Acknowledgments.}-- The authors thank A. Sandvik and A.~J.~Berlinky for useful discussions.
This work was made possible by the
computing facilities of SHARCNET and CESGA.  Support was provided by NSERC
of Canada (A.B.K. and R.G.M.).

\end{document}